\documentclass[%
amsmath,amssymb,
preprint,%
]{revtex4-1}

\usepackage{graphicx}
\usepackage{dcolumn}
\usepackage{bm}

\usepackage[utf8]{inputenc}
\usepackage[T1]{fontenc}
\usepackage{mathptmx}
\usepackage{etoolbox}
\usepackage{svg}
\usepackage{multirow}
\usepackage{amsmath}
\usepackage{subcaption}

\usepackage{graphicx}
\usepackage{float}
\usepackage{subcaption}
\usepackage{hyperref}

\hypersetup{
colorlinks = true,
urlcolor = blue,
citecolor= blue,
}

\makeatletter
\def\@email#1#2{%
 \endgroup
 \patchcmd{\titleblock@produce}
  {\frontmatter@RRAPformat}
  {\frontmatter@RRAPformat{\produce@RRAP{*#1\href{mailto:#2}{#2}}}\frontmatter@RRAPformat}
  {}{}
}%
\makeatother
\begin{document}

\preprint{AIP/123-QED}

\title[Physics of Fluids]{Inertial focusing of neutrally buoyant spherical particle in shallow microchannels\\}
\author{Guiquan Wang}
\author{Willem Van Roy}
\author{Chengxun Liu}
\author{Tim Stakenborg}
\author{Benjamin Jones}

\affiliation{imec, Kapeldreef 75, 3001 Leuven, Belgium}

\date{\today}

\begin{abstract}
This study investigates the lift force acting on a finite-size, neutrally buoyant spherical particle suspended in a liquid while flowing through a shallow channel at low Reynolds numbers. Using an immersed boundary method, we calculate the lift force for particle radius-to-channel height ratios spanning \(0.03 \leq a/H \leq 0.35\) in 2D planar Poiseuille flows. We propose an explicit formula that accurately predicts the lift force for particles as large as \(a/H = 0.35\) and remains valid for particle Reynolds number \(Re_p \leq 1\), despite a reduction in near-wall lift force at higher \(Re_p\). The influence of slip boundary conditions is also explored, demonstrating that increased slip length reduces near-wall lift force and shifts the particle equilibrium position closer to the wall. Predictions of the particle trajectory from the derived model are in good agreement to the published experimental data. These findings offer a practical framework for estimating the migration of large particles in microfluidic devices.\
\end{abstract}

\maketitle

\section{\label{sec:1.Introduction} Introduction }
Precise control of particle motion at the microscale is crucial for harnessing the potential of microfluidics in biological research and healthcare applications\cite{martel2014inertial}.\
One such technique is inertial focusing, which uses inertial forces in microchannel flows to passively position particles into well-defined streamlines. It is particularly attractive due to its simplicity, label‑free operation, broad compatibility, and scalability. \cite{amini2014inertial,xiang2022inertial}.\
The migration of particles during inertial focusing is primarily attributed to the emergence of the lateral lift forces as the inertial aspects of the fluid flow become prominent\cite{di2007continuous}. As a result of the lateral forces, particles traverse across laminar flow streamlines to their equilibrium positions where the lateral forces on the particle are balanced \cite{segre1962behaviourA,segre1962behaviourB}.\
The trajectories of relatively large particles within shallow microchannels (where, typically, $a/H \ge 0.15$ with $a$ is the particle radius and $H$ is the channel height) has important applications such as the separation of T lymphocytes from whole blood\cite{shevkoplyas2005biomimetic}, the separation and evaluation of T lymphocyte activation states by electronic discrimination ($0.18<a/H<0.32$) \cite{han2020continuous}, and the selective cell capture and analysis using shallow antibody-coated microchannels ($0.32<a/H<0.5$)\cite{jang2012selective,AmazonPaper}. Nevertheless, our comprehension of lateral migration in the regime of large particle radius-to-channel height ratios ($ a/H \ge 0.15$) remains limited.\

\begin{figure}[h]
\includegraphics[width=0.49\textwidth]{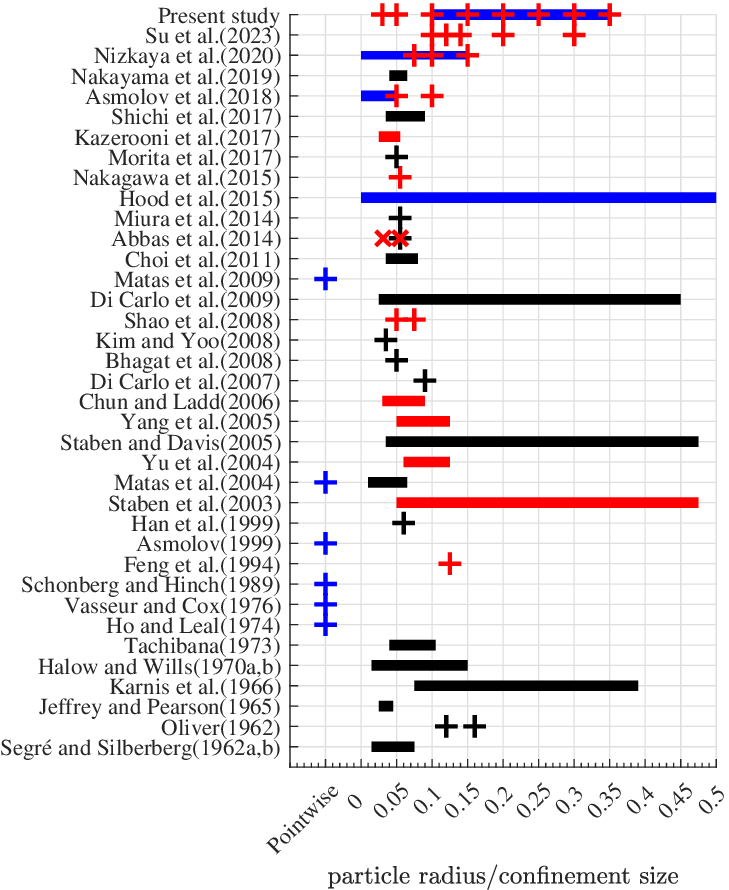}
\caption{\label{fig:Literature_review} Particle migration in the presence of walls as a function of the particle radius-to-confinement size ratio, where confinement size is defined as the channel height or tube diameter. For a comparison of additional parameters, we refer to \citet{shi2020lift}. Numerical simulations (in red), modeling predictions (in blue), and experimental results (in black). Pointwise particle refers to modeling predictions based on the asymptotic approach.}
\end{figure}

Fig.\ref{fig:Literature_review} shows relevant studies of the particle migration during Poiseuille flow; most previous studies have primarily focused on cases where $a/H \le 0.125$ \cite{segre1962behaviourA, segre1962behaviourB,oliver1962influence,jeffrey1965particle,tachibana1973behaviour,ho1974inertial,vasseur1976lateral,schonberg1989inertial,feng1994direct,asmolov1999inertial,han1999particle,matas2004inertial,yu2004dynamic,yang2005migration,chun2006inertial,di2007continuous,bhagat2008enhanced,kim2008lateral,shao2008inertial,matas2009lateral,choi2011lateral,abbas2014migration,miura2014inertial,nakagawa2015inertial,morita2017equilibrium,kazerooni2017inertial,shichi2017inertial,asmolov2018inertial,nakayama2019three,nizkaya2020inertial,su2023new}. However, there is a paucity of research investigating cases where $a/H > 0.15$ \cite{oliver1962influence, Karnis1966, staben2003motion, staben2005particle, di2009particle, hood2015inertial,su2023new}, particularly when the particle diameter exceeds half the channel height $a/H \ge 0.25$.\
Among these, \citet{oliver1962influence} demonstrates the importance of particle rotation on the migration direction, showing that a rotating sphere moves outward, whereas a non-rotating sphere tends to drift toward the tube center;\ \citet{Karnis1966} experimentally finds the particle is force-free at $z_{eq} = 0.25 H$ ($z_{eq}$ is the equilibrium location from the particle center to the tube wall) for $a/H = 0.15$ and shifts towards the tube cente as $a/H$ increases to $0.39$;\ 
\citet{staben2003motion} observed in numerical simulations that particle streamwise slip velocity increases with the particle to channel size ratio, and \citet{staben2005particle} experimentally validated these findings. However, these studies did not provide lift force;\
\citet{di2009particle} experimentally finds the particle force-free location at $z_{eq}/H \sim 0.3$ for $a/H = 0.05$ and shifts towards the center as $a/H$ increases to $0.45$ in a square duct. Additionally, through numerical simulations, they find the lift-force is proportional to $a^3$ far from the wall while it is proportional to $a^6$ in the near-wall region for $0.22 \le a/H \le 0.38$. \
To clarify the size dependence of inertial lift forces acting on particles in microchannels, \citet{su2023new} decomposes the lift force into pressure-induced and viscosity-induced components, analyzing their scaling separately. In contrast, \citet{hood2015inertial} provides a more physically grounded explanation by theoretically deriving a model incorporating components depending on both an \( a^4 \) term and a correction term \( a^5 \). This model successfully demonstrates that the scaling observed by \citet{di2009particle} arises from an incidental fitting of a perturbation series in \( a \) to a single apparent scaling law.\

Predicting the lift force is challenging. In a simplified case of a neutrally buoyant spherical particle in a pressure-driven microfluidic channel flow, the lateral force arises from several contributions: wall repulsion due to lubrication, inertial lift associated with shear slip, lift due to particle rotation, and lift induced by the curvature of the undisturbed velocity profile \cite{feng1994direct}.\ 
Experimentally measuring the lateral lift force is particularly difficult. As a result, theoretical approaches have been extensively pursued over the past several decades. For small neutrally buoyant particles (\(a/H \ll 1\)) at low channel Reynolds numbers (\(Re \ll 1\)), \citet{ho1974inertial} used an asymptotic approach to predict the lift force on freely rotating circular particles in a 2D geometry. Their method assumes that the nearest wall lies within the particle’s inner region, where inertial terms are negligible compared to viscous terms. They applied Lorentz’s method of reflection to solve the perturbation equations. Later, \citet{vasseur1976lateral} analyzed the migration velocity of a spherical particle in planar Poiseuille flow, predicting the lift force using a point-force approximation. However, both methods agree well only when the particle is far from the wall (\(z_p/a \gg 1\), where \(z_p\) is the particle’s wall-normal position from the wall).\
For \(z_p/a \sim 1\), \citet{cherukat1994inertial} introduced a simple fitted correlation for a particle near a single wall in linear shear flow, based on numerical results for finite-size particles.\

For finite-size particle at any wall-normal location, by a speculate approach, \citet{asmolov2018inertial} combine the coefficients obtained by \citet{vasseur1976lateral} and \citet{cherukat1994inertial} to generalized expressions for the lift force of small particles ($a/H \le 0.05$) and finite $Re \le 20$. For larger particles ($a/H \le 0.15$), \citet{nizkaya2020inertial} propose a modification of expressions for the coefficients reported by \citet{asmolov2018inertial} by fitting to their own simulation data. The key advantage of the models proposed by \citet{asmolov2018inertial} and \citet{nizkaya2020inertial} is their simplicity and accuracy in predicting the lift force, making them well-suited for quick design in microfluidic applications.\
For very large finite-size particles (\(a/H > 0.15\)) at low particle Reynolds number (\(Re_p \ll 1\)), \citet{hood2015inertial} numerically investigated the dominance of viscous stresses over inertial stresses across a broader range of \(Re\) and particle sizes, extending the theory of \citet{ho1974inertial} to spherical particles of any size in Poiseuille flow. Remarkably, the coefficients in the lift force expression depend solely on the particle's location, independent of its size. However, the complex expressions for the coefficients must be determined through numerical simulations, limiting the model’s direct applicability in microfluidic device design.\

In this paper, we introduce a simple and practical method to predict the coefficients of \(a^4\) and \(a^5\) where $a$ denotes the particle radius. To validate the method, we conduct a numerical investigation of finite-size, neutrally buoyant spherical particles (\(a/H \leq 0.35\)) in the low particle Reynolds number regime (\(Re_p \leq 1\)), with a focus on lift forces and equilibrium positions. The paper is structured as follows: Section~\ref{sec:2.Methodology} details the proposed model formulation and numerical simulation methodology, including grid independence analysis and validation against published benchmark studies. Section~\ref{sec:3.Results} validates the proposed model of lift force using extensive simulation results. Section~\ref{sec:4.Conclusion} summarizes key findings and discusses application of the model to predicting particle trajectories. Appendix ~\ref{app:LiftModel} provides the lift force coefficients in prior studies and Appendix ~\ref{app:Stokes_correction} validate Stokes drag corrections for migration velocity predictions.  \\

\section{\label{sec:2.Methodology} Methodology}

We consider a sphere with radius $a$ in a laminar pressure-driven flow between two parallel walls with a distance $H$, representing a 2D planar Poiseuille flow. {This represents the depth‑averaged limit of a shallow microchannel whose spanwise dimension is much larger than its height. In this regime, the dominant confinement length scale governing inertial lift is the channel height, justifying the use of the particle size ratio $a/H$. The model neglects spanwise velocity gradients and secondary flows, allowing isolation of wall‑induced and shear‑induced inertial lift mechanisms in a controlled planar setting.} The flow is driven by a pressure gradient, resulting in a parabolic velocity profile for the undisturbed flow, $U(z)=4U_m z/H (1-z/H)$, where $U(z)$ is the streamwise velocity, $U_m$ is the velocity at the central plane, $z$ is the wall-normal distance from the bottom wall. The channel Reynolds number is defined as $Re=\rho \left< U \right> H / \mu$ where $\rho$ is the fluid density, {$H$ is the wall-to-wall channel height in 2D planar channel flow},$\left< U \right>$ is the average channel flow velocity and $\mu$ is the dynamic viscosity. The sphere's position in the wall-normal direction is given by its center $z_p$ from the bottom wall. The particle Reynolds number is defined as $Re_p=(a/H)^2 Re$. In this study, we focus on neutrally buoyant particles, meaning the particle density is equal to the fluid density.

\subsection{\label{sec:2.0 Model} Model}

For inertial migration of neutrally buoyant finite-sized particles in microchannel fluid flow, the lift force \cite{asmolov1999inertial} is defined as 

\begin{equation}
F_{L}=\rho a^4 G_m^2 c_L,
\label{eq:Fl_asmolov}
\end{equation}

\noindent where $G_m=4U_m/H$ is the maximum shear rate at the wall. 
For small particle size $a/H \le 0.05$, \citet{asmolov2018inertial} proposed one of the first models with explicit formula of the lift coefficient $c_{L}$ which is given in Eq.(\ref{eq:cl_asmolov}). For medium particle size $a/H \le 0.15$, \citet{nizkaya2020inertial} proposed a modification of expression of \citet{asmolov2018inertial} based on simulation data, the corrected lift coefficient is given in Eq.(\ref{eq:cl_Nizkaya}). For any particle size $0< a/H < 0.5$, \citet{hood2015inertial} develop a perturbation series expansion for the lift force coefficient with the assumption that the wall lying in the inner region perturbed by the particle. The expression reads off

\begin{equation}
c_{L}=c_{4} + c_{5} \frac{a}{H},
\label{eq:cl_Hood}
\end{equation}
\noindent where the coefficients $c_{4} $ and $c_{5}$ in the expression depend only on the location of the particle regardless of the particle size. However, the coefficients need to be calculated from direct numerical simulations.

Here we propose a simple explicit formula for $c_{4} $ and $c_{5}$ in the expression of Eq.(\ref{eq:cl_Hood}) based on the lift coefficients expression in Eq.(\ref{eq:cl_Nizkaya}) of \citet{nizkaya2020inertial} from two particle sizes ($a_1$ and $a_2$)

\begin{eqnarray}
c_{4}(a/H)&=&\frac{(H/a_2) \cdot c_{L}(a_2/H)- (H/a_1) \cdot c_{L}(a_1/H)}{H/a_2-H/a_1},\nonumber\\
c_{5}(a/H)&=&\frac{c_{L}(a_2/H)-c_{L}(a_1/H)}{a_2/H-a_1/H}, \nonumber\\
\label{eq:cl_present}
\end{eqnarray}

\noindent where $c_{L}(a/H)$ as a function of the normalized particle radius $a/H$, evaluated at $c_{L}(a_1/H)$ and $c_{L}(a_2/H)$ calculated by Eq.(\ref{eq:cl_Nizkaya}), $a_1$ and $a_2$ are in the range of $a_1/H \le 0.15$ and $a_2/H \le 0.15$ within the validated range of Eq.(\ref{eq:cl_Nizkaya}). 

{The mixed scaling form in Eq.(\ref{eq:cl_Hood}) follows the perturbation framework of \citet{hood2015inertial}, in which the lift coefficient is expressed as a sum of $a^4$ and $a^5$ contributions with coefficients depending only on particle position. In the present work, the coefficients $c_4$ and $c_5$ are obtained using a simple interpolation strategy based on two reference particle sizes, $a_1$ and $a_2$, within the range where existing models are well validated.
The parameters $a_1$ and $a_2$ therefore serve as practical fitting anchors, rather than quantities with independent physical meaning. Their values are selected to optimally capture different confinement regimes across particle sizes, balancing near‑wall and finite‑size effects, and are not intended to represent distinct physical transitions in the migration mechanism.}
By comparing our simulation data for selected $ 0.1 \leq a/H \leq 0.35$, we suggest three groups of $a_1$ and $a_2$ as shown in Table~\ref{tab:a1a2_select}. We note that the selected values of $a_1$ and $a_2$ are based on a limited set of simulated $a$ values and may not be optimal for all possible $a$. Additionally, our tests were conducted at as large particle size as $a/H=0.35$, without exploring the upper limits of this model.

\begin{table}[h]
\caption{Suggested values for \( a_1 \) and \( a_2 \) of $c_{L}$ for predicting lift coefficients $c_4(a/H)$ and $c_5(a/H)$ in Eq.~(\ref{eq:cl_present}) for a given \( a \). {The values of \( a_1 \) and \( a_2 \) are chosen as fitting anchors to interpolate the mixed‑scaling lift coefficient across different particle‑size regimes; they do not correspond to intrinsic physical length scales.}}
\label{tab:a1a2_select}
\begin{ruledtabular}
\begin{tabular}{c c c}
$a/H$                & $a_1/H$ & $a_2/H$ \\
\hline
$a/H<0.2$            & 0.01 & 0.05 \\
$0.2 \leq a/H <0.35$ & 0.05 & 0.15 \\
$a/H\geq 0.35$       & 0.01 & 0.15 \\
\end{tabular}
\end{ruledtabular}
\end{table}


\subsection{\label{sec:2.1 method}Numerical method and mesh}

Direct numerical simulations of single-phase flows are performed using the AFiD\cite{van2015pencil} code for an incompressible Newtonian fluid. The flow is periodic in the two horizontal directions and wall-bounded in the vertical direction.\
Spatial discretization employs a conservative second-order centered finite difference scheme with velocities defined on a staggered grid, while pressure is computed at the cell center. The nonlinear terms are discretized using the explicit Adams-Bashforth scheme, and the viscous terms are treated with the implicit Crank-Nicholson scheme. Time integration is performed using a fractional-step third-order Runge-Kutta (RK3) method.\
For the dispersed phase, numerical simulations are based on the moving least squares immersed boundary method (IBM), where the particle interface is represented by a triangulated Lagrangian mesh \cite{de2016moving, spandan2018fast}.\  

\begin{figure}[h]
\centering
\includegraphics[width=0.38\textwidth]{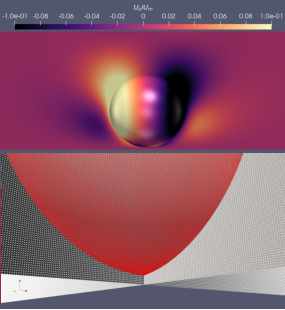}
\caption{\label{fig:Contour_mesh} The top panel shows a contour plot of the wall-normal velocity, normalized by the centerline velocity, on the cross-section of the spanwise direction for the fluid and on the particle surface, for \(a/H = 0.3\) and \(z_p/H = 0.32\) at \(Re = 1\) with \(N_g \simeq 6\). The bottom panel illustrates the triangular Lagrangian mesh on the particle (in red) and the Eulerian mesh on two cross-sections (in gray).}
\end{figure}

The Navier–Stokes equations for an incompressible Newtonian flow, including a body force \(\mathbf{f_b}\) applied in the vicinity of a solid particle, are expressed as follows:
\begin{eqnarray}
\nabla \cdot \mathbf{u}&=&0,\\
\rho_f \frac{D\mathbf{u}}{D t}&=&-\nabla p + \mu \nabla^2 \mathbf{u} + \mathbf{f_{b}}(\mathbf{x}, t), 
\end{eqnarray}
where $\mathbf{u}$ is the velocity, $p$ is the pressure, $\rho_f$ is the fluid density and $\mu$ is the dynamic viscosity. The momentum equation is solved throughout the entire domain, including the regions occupied by the particles. $\mathbf{f}_{b}$ is introduced to enforce the interfacial no-slip boundary condition. The interfacial boundary condition is enforced by first computing the required force at the Lagrangian markers and then transferring it to the Eulerian mesh \cite{uhlmann2005immersed}. The details of the methodology to construct the Lagrangian mesh and the schemes used to transfer flow quantities between the Lagrangian and Eulerian mesh are reported in \citet{de2016moving,spandan2018fast}. The governing equations for the solid particles are advanced every RK time step\cite{breugem2012second}. The ratio of the mean edge length of Lagrangian mesh to the Eulerian mesh uniform spacing is set between $0.8$ and $0.9$, a range shown to ensure smooth flow interactions between the immersed body and the fluid \cite{spandan2018fast, breugem2012second}. The time step is chosen according to \citet{breugem2012second},
\begin{equation}
    \Delta t \le min \left(  \frac{1.65}{12} \frac{\Delta x^2}{\nu}, \frac{\sqrt{3} \Delta x}{\sum^3_{i=1}| {u_i}|}\right),
\end{equation}
where $\nu$ is the kinetic viscosity and $\Delta x$ is the mesh size for a uniform Eulerian grid. In Fig.~\ref{fig:Contour_mesh}, the top panel shows the wall-normal velocity contour of a sphere with \( a/H = 0.3 \) in a flow at \( Re = 1 \) located at \( z_p/H = 0.32 \). The bottom panel displays the grid points of both the Lagrangian markers (with 327,680 faces on the sphere's surface) and the Eulerian mesh (321 regular grid points in the wall-normal direction). The flow field remains smooth when the mean edge length of the triangular Lagrangian markers is smaller than the local Eulerian grid spacing, with \( N_g \simeq 6 \) grid points in the narrowest gap between the sphere and the wall. {Here, $N_g$ denotes the number of Eulerian grid points resolving the minimum particle–wall gap. Because inertial lift is most sensitive to near‑wall interactions, convergence with respect to $N_g$ directly assesses the adequacy of gap resolution, while the Eulerian grid is refined uniformly over the entire domain}.

The lift force is the integration of the pressure and viscous stresses over the particle surface
\begin{equation}
\bm{F}_L=\int_{\sum_p} \left( -p \bm{I} + \bm{\tau} \right) \cdot \bm{n} ds,
\end{equation}
where $\sum_p$ represents the particle surface, \(ds\) denotes the triangulated Lagrangian mesh, \(\bm{I}\) is the unit tensor, \(\bm{\tau}\) is the shear rate tensor, and \(\bm{n}\) is the unit normal vector to the particle surface. However, the particle will not reach a steady state until it attains its equilibrium position. Instead, we apply a force opposite to the particle transverse velocity to prohibit the particle wall-normal motion which theoretical predicts the quasi-steady lift force \cite{gupta2018conditional,asmolov2018inertial}. The applied force is set to zero initially, then updated at iteration $k$ from the value at iteration $k-1$, according to a penalty method $F_{ext}(k)=F_{ext}(k-1)-\Gamma[6\pi\mu a W_p(k-1)]$ proposed by \citet{gupta2018conditional} where $W_p(k-1)$ represent for the particle lateral migration velocity from the previous time step in the wall-normal direction. The iteration stops until the particle transverse velocity is very close to zero. $\Gamma$ is an arbitrary constant which should be chosen not very low in order to reduce the time needed for convergence and not very high in order to avoid numerical instability. For $Re=1$, a reference value $\Gamma=0.02$ is given which is tested to be suitable to all size of particles used in this study. \

\begin{figure}[h]
\centering
\includegraphics[trim={0cm 0cm 9cm 0cm}, clip,width=0.38\textwidth]{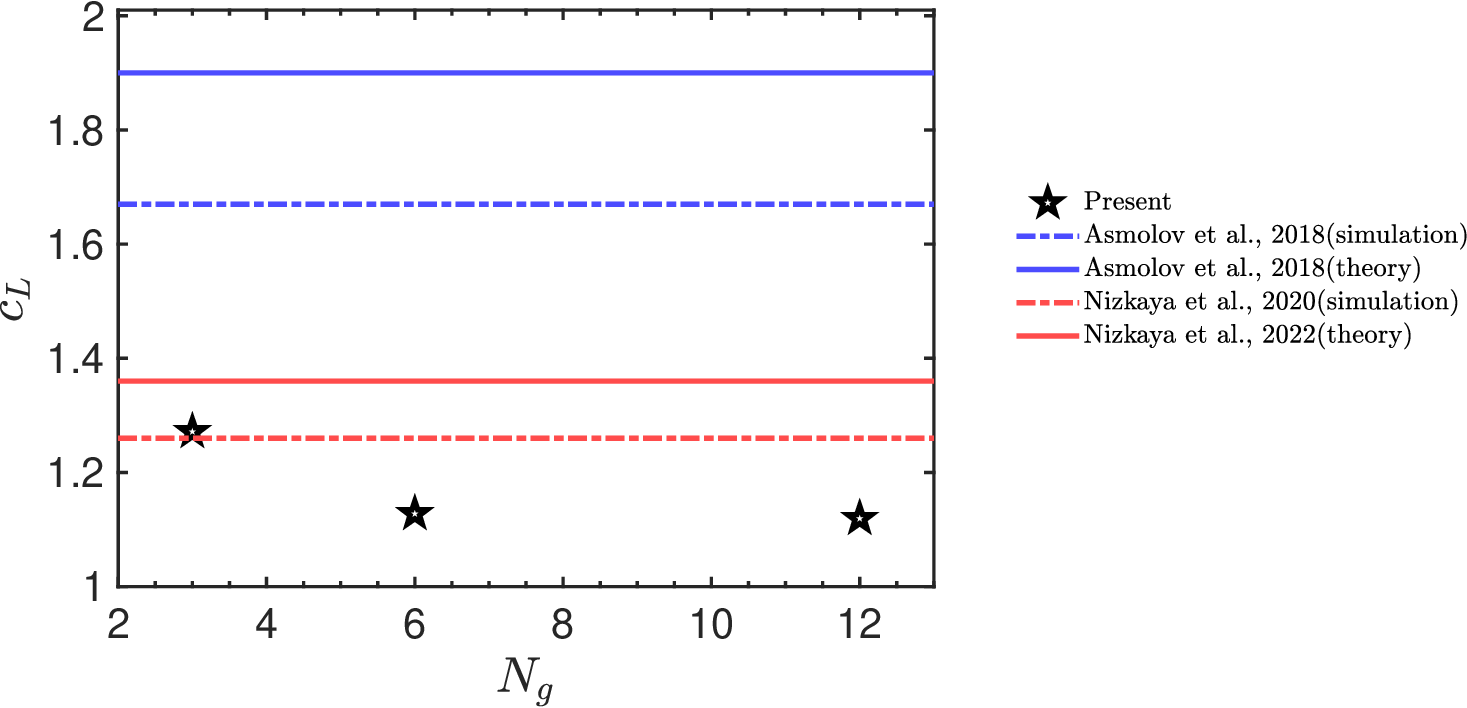}
\caption{\label{fig:cL_Ng_a01} {Mesh‑convergence study of the lift coefficient $c_L$ with respect to $N_g$, the number of Eulerian grid points in the narrowest particle–wall gap, for $a/H=0.1$, $z_p/H=0.11$, and $Re=1$}.Present results (pentagrams). Horizontal lines represent literature values, from top to bottom: \citet{asmolov2018inertial} model prediction (blue solid line) and LBM results (blue dash-dotted line), \citet{nizkaya2020inertial} model prediction (red solid line) and LBM results (red dash-dotted line). {Literature results are shown for comparison only and do not represent mesh‑refined data within the present study.}
}
\end{figure}

To test $N_g$ effect on the lift force, we perform the $N_g$ independent test when the particle is in the vicinity of the wall. In order to compare with the previous published data \cite{asmolov2018inertial, nizkaya2020inertial}, the simulations are set for particle size $a/H=0.1$ and channel Reynolds number $Re=1$ at $z_p/H=0.11$. As shown in Fig.~\ref{fig:cL_Ng_a01}, the present result of $N_g=3$ agrees well with simulation result of \citet{nizkaya2020inertial} whereas about $30$ percents smaller than simulation result of \citet{asmolov2018inertial}. We note that the lattice Boltzmann method (LBM) is employed in both \citet{asmolov2018inertial} and \citet{nizkaya2020inertial}, in which there are about $80$ lattices placed in the wall normal direction corresponding to $1\sim 2$ lattices in the nearest gap between the sphere and wall. When comparing with the model predictions, \citet{nizkaya2020inertial} significantly improves the predicted value than \citet{asmolov2018inertial}. However, increasing $N_g$ from $3$ to $6$, the lift force coefficient $c_L$ shows strong dependency on $N_g$, which decreases about $13$ percents. Further increasing $N_g$ from $6$ to $12$, $c_L$ decreases less than $3$ percents. In this work, for most cases we place at least $N_g \ge 12$ Eulerian grids. However, taking the computational cost into account, in several computationally expensive cases when particle is in the vicinity of the wall, we place $N_g = 6$ or $7$ Eulerian grids to calculate $c_L$. Additionally, the domain size in the streamwise and spanwise directions is generally set to \( L_x = 20a \) and \( L_y = 10a \) to ensure domain independence, respectively. In scenarios where the particle is very close to the wall, we reduce the streamwise domain size to \( L_x = 10a \) to reduce computational cost. \\ 

\subsection{\label{sec:2.2 validation}Validation of the numerical method}

In Poiseuille flow, the lift force arises from the sphere's interaction with complex flow phenomena. When the particle is near the central region of the channel, the interaction between the particle's stresslet and the parabolic velocity profile generates a lift force directed toward the wall \cite{ho1974inertial}. Conversely, when the particle is close to the wall, the slip velocity of the particle in the presence of shear induces a lift force directed toward the channel center \cite{cherukat1994inertial}. Between the channel center and the walls, the particle traverses streamlines until it reaches a force-free equilibrium position. Accurately capturing these effects requires a well-implemented numerical scheme, making the lift force a robust criterion for validating numerical codes. \\

\begin{figure}
    \centering
    \begin{minipage}{0.48\linewidth}
        \centering
        \includegraphics[trim={0cm 0cm 7.5cm 0cm}, clip,width=1\textwidth]{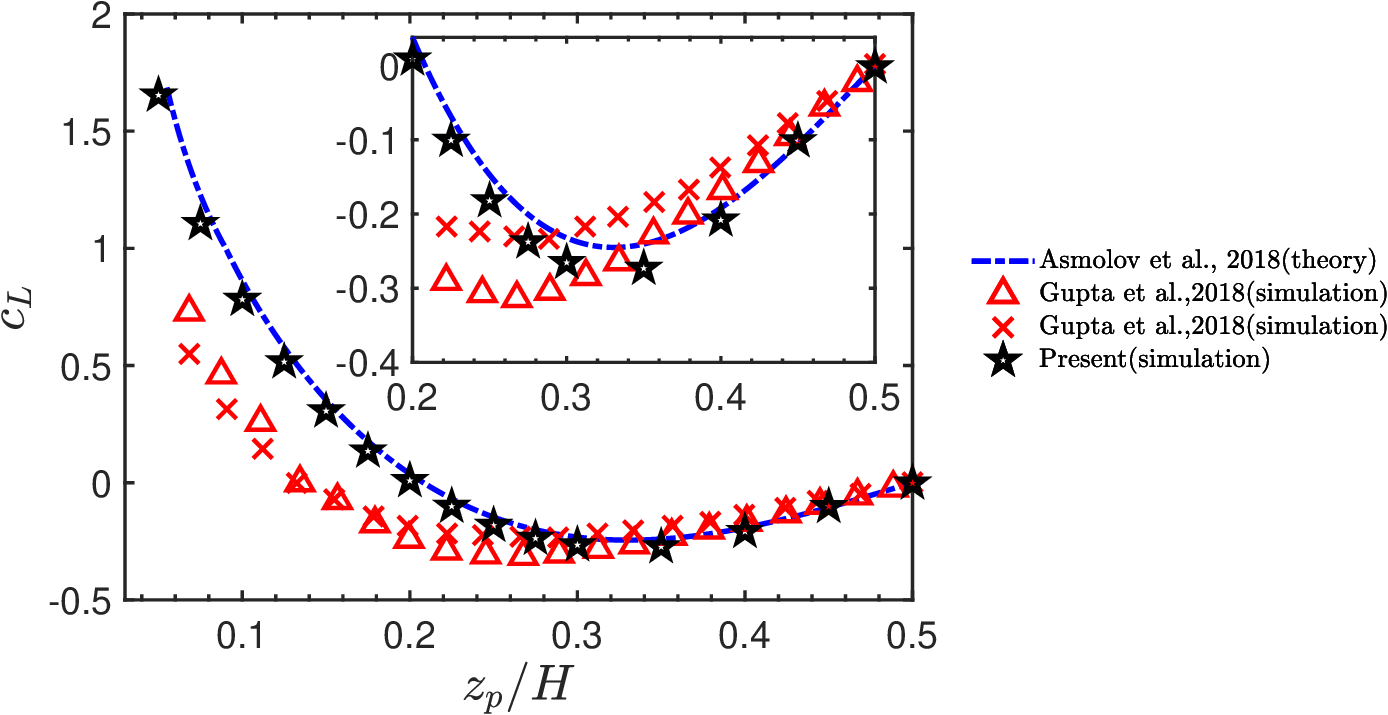}
        \caption*{(a)}
        \label{fig:cL_aH003}
    \end{minipage}
    \hfill
    \begin{minipage}{0.49\linewidth}
        \centering
        \includegraphics[trim={0cm 0cm 8cm 0cm}, clip,width=1\textwidth]{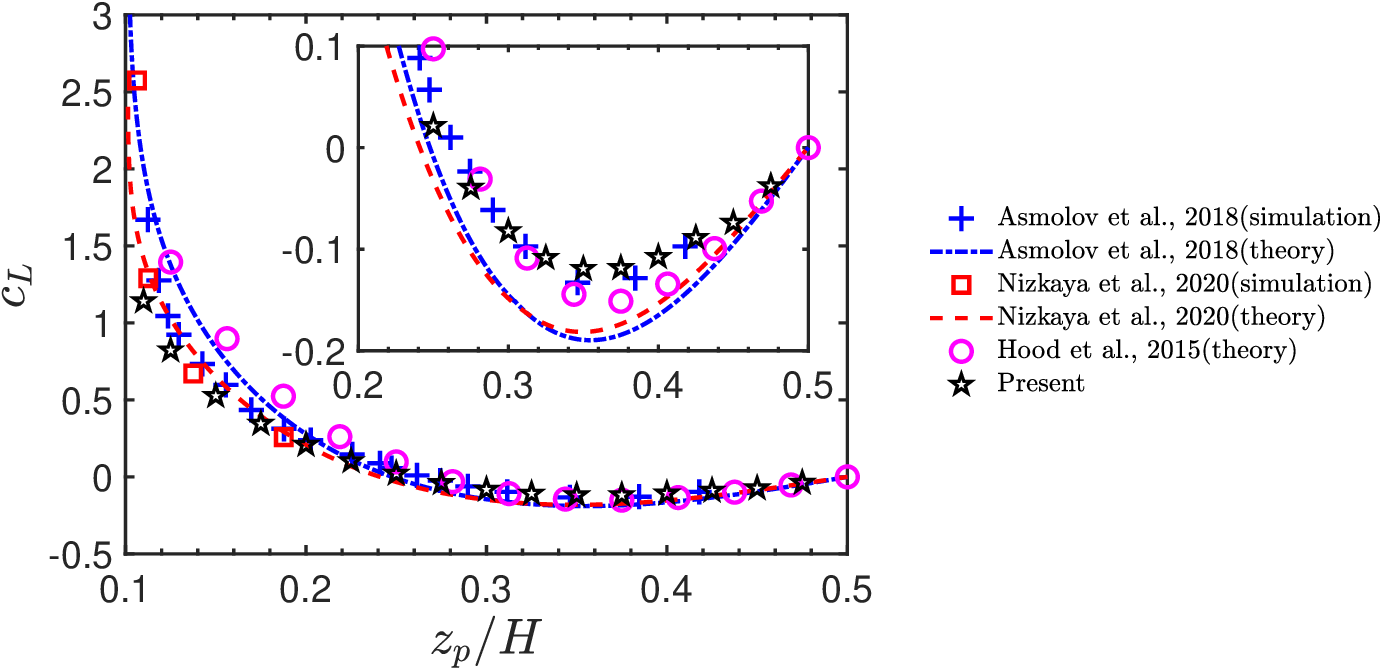}
        \caption*{(b)}
        \label{fig:cL_aH01}
    \end{minipage}

\caption{Lift coefficient \(c_L\) for two particle sizes: (a) \(a/H = 0.03\), present results at \(Re = 1\) (black pentagrams), model prediction by \citet{asmolov2018inertial} (blue dash-dotted line), with triangles and crosses representing results from \citet{gupta2018conditional} at \(Re = 13\) and \(Re = 38\), respectively; (b) \(a/H = 0.1\), present results at \(Re = 1\) (black pentagrams), model predictions by \citet{asmolov2018inertial} (blue dash-dotted curve), \citet{nizkaya2020inertial} (red dashed curve), and \citet{hood2015inertial} (magenta circles), along with LBM results from \citet{asmolov2018inertial} (blue plus symbols) and \citet{nizkaya2020inertial} (red squares). {Insets provide magnified views highlighting local differences between numerical and model results, particularly where $c_L$ is small.}}
\label{fig:cL_validation}
\end{figure}

The lift force for two particle sizes (\(a/H = 0.03, 0.1\)) at \(Re = 1\) across different channel locations (\(z_p\)) is obtained from numerical simulations. 
For small particles (\(a/H = 0.03\)), as shown in Fig.~\ref{fig:cL_validation}(a), {the present results align well with the model predictions of \citet{asmolov2018inertial}, which are theoretically valid for \(a/H \ll 1\) and \(Re_p \ll 1\), while remaining applicable for channel Reynolds number up to $Re \lesssim 20$}. Compared to the simulation results of \citet{gupta2018conditional} at \(Re = 13\) and \(38\), the present lift force is smaller near the wall (\(z_p/H < 0.3\)) but becomes higher toward the channel center (\(0.3 < z_p/H < 0.5\)). Additionally, the results indicate a stronger \(Re\)-dependence of the lift force near the wall compared to the channel center. 
The observed differences between the present data and published results may stem from two factors. First, the numerical schemes differ: the force-coupling method\cite{wang2017modulation} used by \citet{gupta2018conditional} represents the particle’s finite size using two Gaussian spherical envelopes for the monopole and dipole, without enforcing a no-slip boundary on the rigid sphere. This could lead to variations in the effective particle size, rotation rate, and slip velocity compared to the surface-tracking IBM numerical scheme used here. Second, the \(Re\) effect plays a role; the inertial interaction with the wall decreases as \(Re\) increases \cite{schonberg1989inertial}, resulting in a reduced lift force, as also reported by \citet{gupta2018conditional}.

For medium-sized particles (\(a/H = 0.1\)), as shown in Fig.~\ref{fig:cL_validation}(b), the present results show good agreement with the simulation data from \citet{asmolov2018inertial} and \citet{nizkaya2020inertial} for \(z_p/H > 0.13\). However, very close to the wall (\(z_p/H < 0.13\)), the published simulation results are higher than the present findings. This discrepancy may be attributed to insufficient lattice resolution in the narrow gap between the sphere and the wall, as depicted in Fig.~\ref{fig:cL_Ng_a01}.
The numerical results are also compared with the prior model predictions\cite{asmolov2018inertial,nizkaya2020inertial,hood2015inertial}, with the applicability conditions of these models detailed in Section~\ref{app:LiftModel}. Near the wall (\(z_p/H < 0.25\)), the predictions from \citet{asmolov2018inertial} and \citet{hood2015inertial} are higher than the numerical results, whereas \citet{nizkaya2020inertial} demonstrates significantly improved agreement. Toward the channel center (\(0.25 \leq z_p/H \leq 0.5\)), as shown in the inset of Fig.~\ref{fig:cL_validation}(b), the predictions from \citet{hood2015inertial} align better with the numerical results compared to those from \citet{asmolov2018inertial} and \citet{nizkaya2020inertial}.  
Across the channel, from the wall to the center, it is observed that none of the three model predictions consistently outperforms the others.\\




      
      
      

\section{\label{sec:3.Results} Results and discussion}

\subsection{\label{sec:3.1 particle size} Scaling law and model prediction for lift Force}

Previous studies have demonstrated that the lift force on a sphere in Poiseuille flow scales with the particle radius, but the scaling exponents vary depending on the particle size and its position within the channel. For small particles in a channel, \citet{ho1974inertial} first theoretically derived the lift force as \(F_L \sim c_L a^4/H^2\), showing that the lift force scales with a power law of \(a^4\), under the assumptions \(a \ll 1\) and \(Re \ll 1\). Subsequently, \citet{schonberg1989inertial} and \citet{asmolov1999inertial} extended the parameter space to higher Reynolds numbers (\(Re = O(1)\) and \(Re = O(10^3)\), respectively), and both studies retained the same \(a^4\) scaling as in \citet{ho1974inertial}.\

\begin{figure}
\centering
\begin{subfigure}{0.38\textwidth}
\includegraphics[trim={0cm 0cm 0cm 0cm}, clip,width=1\textwidth]{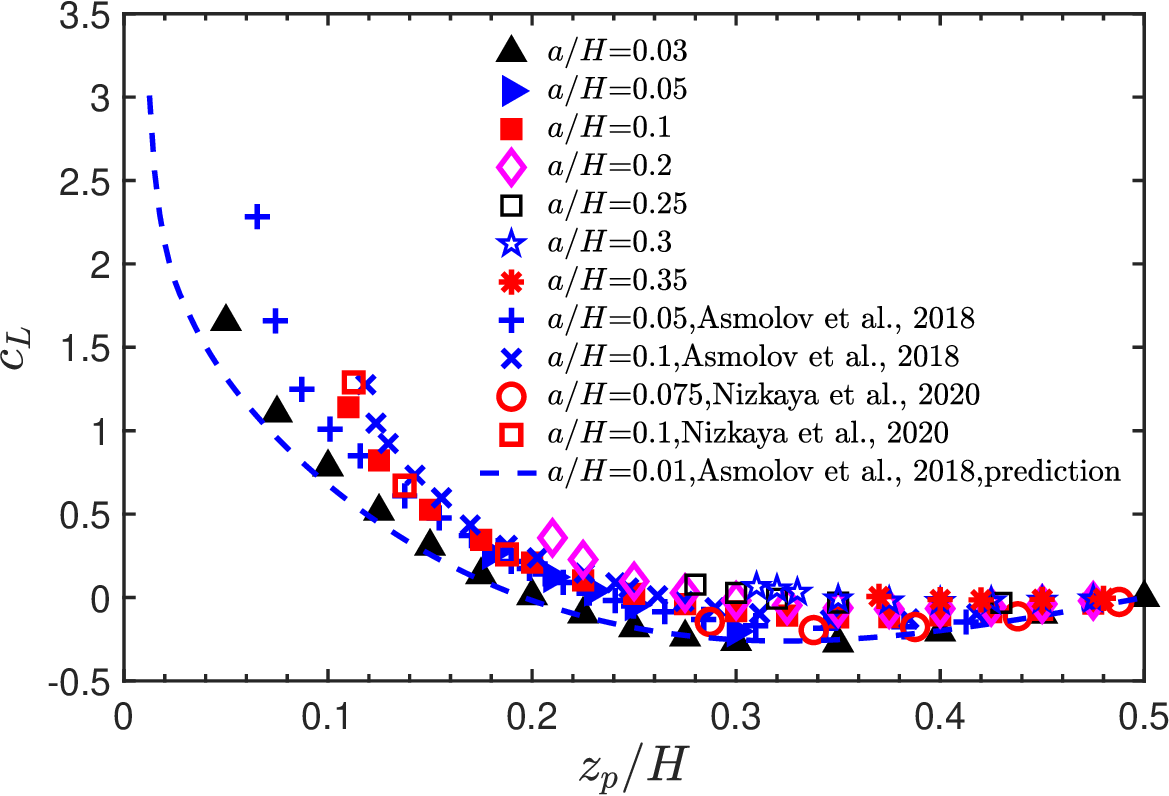}
\caption{}
\label{fig:cL_a4}
\end{subfigure}

\begin{subfigure}{0.38\textwidth}
\includegraphics[trim={0cm 0cm 0cm 0cm}, clip,width=1\textwidth]{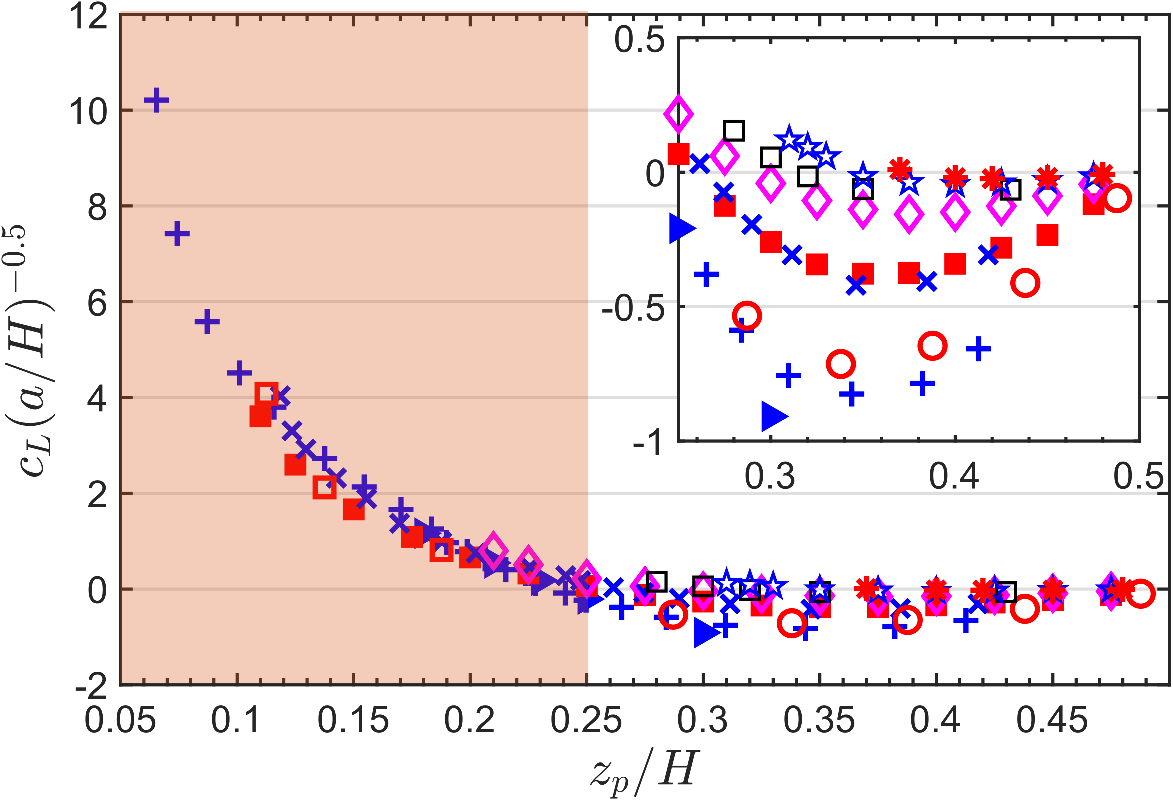}
\caption{}
\label{fig:cL_a4p5}
\end{subfigure}

\begin{subfigure}{0.38\textwidth}
\includegraphics[trim={0cm 0cm 0cm 0cm}, clip,width=1\textwidth]{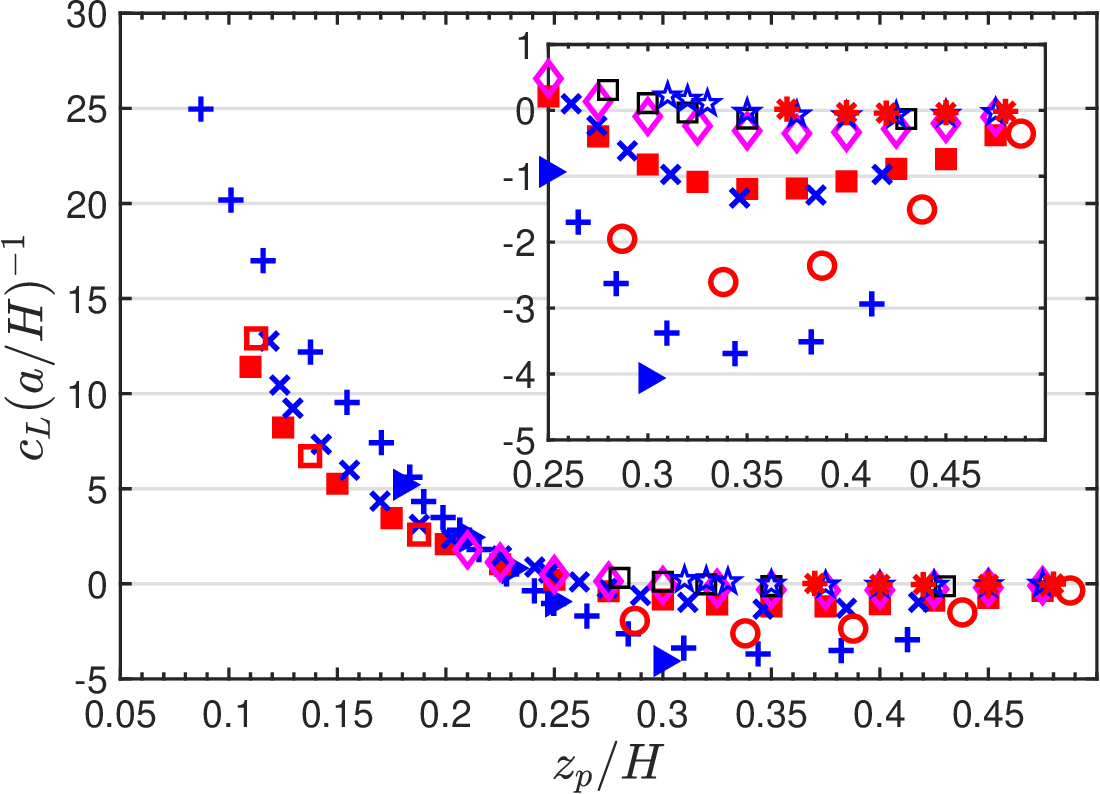}
\caption{}
\label{fig:cL_a5}
\end{subfigure}
\caption{Lift force under different scaling laws for particle radius-to-channel height ratio \(0.01 \leq a/H \leq 0.35\): (a) \(c_L\) representing the \(F_L\) scaling law with \(a^{4}\); (b) \(c_L\) plotted with a new functional dependence on \((a/H)^{-0.5}\), representing the \(F_L\) scaling law with \(a^{4.5}\), which better collapses the data near the wall; (c) \(c_L\) plotted with a functional dependence on \((a/H)^{-1}\), representing the \(F_L\) scaling law with \(a^{5}\), showing that the data cannot be effectively scaled. {Insets provide magnified views highlighting local differences between numerical results, particularly where $c_L$ is small.}}
\end{figure}

Fig.~\ref{fig:cL_a4} presents $c_L$ representing the \( a^4 \) scaling of the lift force $F_L$, based on numerical simulations for particle sizes ranging from \( a/H = 0.03 \) to \( 0.35 \) and the model prediction for small particles (\( a/H = 0.01 \)) from \citet{asmolov2018inertial}. When considering a broader range of particle sizes, \( F_L \) does not consistently follow the \( a^4 \) power-law scaling, both in the near the wall region and channel center. However, for small particles (\( a/H \leq 0.03 \)), \( F_L \) adheres to the \( a^4 \) scaling. This plot supports the \( a^4 \) scaling law predicted by previous theoretical studies \cite{ho1974inertial, schonberg1989inertial, asmolov1999inertial, hood2015inertial}, under the assumptions for asymptotically small particles. \

For larger particles in a square channel, \citet{di2009particle} observed that when a particle is near the channel center, the lift force follows a power-law scaling of \( a^3 \) instead of \( a^4 \). Closer to the channel wall, the scaling exponent appears to follow \( a^6 \), casting doubt on the validity of earlier asymptotic theories for larger particles. \

Fig.~\ref{fig:cL_a4p5} illustrates the \( a^{4.5} \) scaling of the lift force for particle sizes ranging from \( a/H = 0.05 \) to \( 0.35 \). For \( z_p/H \leq 0.25 \), \( F_L \) adheres to the \( a^{4.5} \) power-law scaling across larger particle sizes (\( 0.05 \leq a/H \leq 0.2 \)). However, as particles approach the channel center, \( F_L \) increasingly deviates from this scaling. Tests of alternative scaling laws (\( a^3 \), \( a^{3.5} \), \( a^5 \), \( a^{5.5} \), and \( a^6 \)) confirm that none consistently describe the data; for brevity, only the \( a^5 \) scaling is shown for comparison in Fig.~\ref{fig:cL_a5}. The physical origin of the \( a^{4.5} \) scaling remains unclear. This behavior, observed for finite-size particles near the wall, contrasts with both the asymptotic theory for small particles \cite{ho1974inertial,schonberg1989inertial} and the \( a^6 \) scaling reported for finite-size particles in confined square channels \cite{di2009particle}. Instead, it aligns with the mixed \( a^4 \)-\( a^5 \) scaling proposed by \citet{hood2015inertial}, who interpreted variations in scaling across channel regions as a coincidental fit of a perturbation series in \( a \) into a single apparent power law. \

\begin{figure}
    \centering
    \begin{minipage}{0.48\linewidth}
        \centering
        \includegraphics[trim={0cm 0cm 0cm 0cm}, clip,width=1\textwidth]{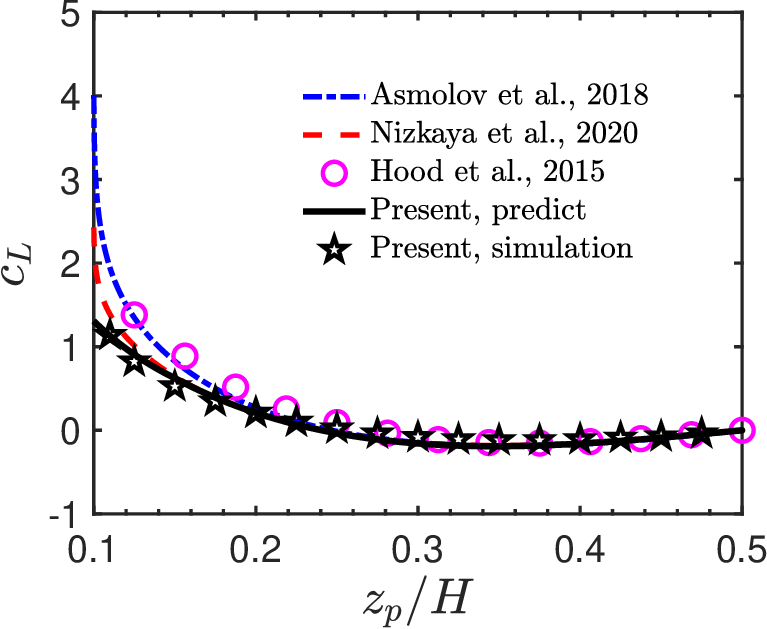}
        \caption*{(a)}
        \label{fig:cL_aH01_model}
    \end{minipage}
    \begin{minipage}{0.49\linewidth}
        \centering
        \includegraphics[trim={0cm 0cm 0cm 0cm}, clip,width=1\textwidth]{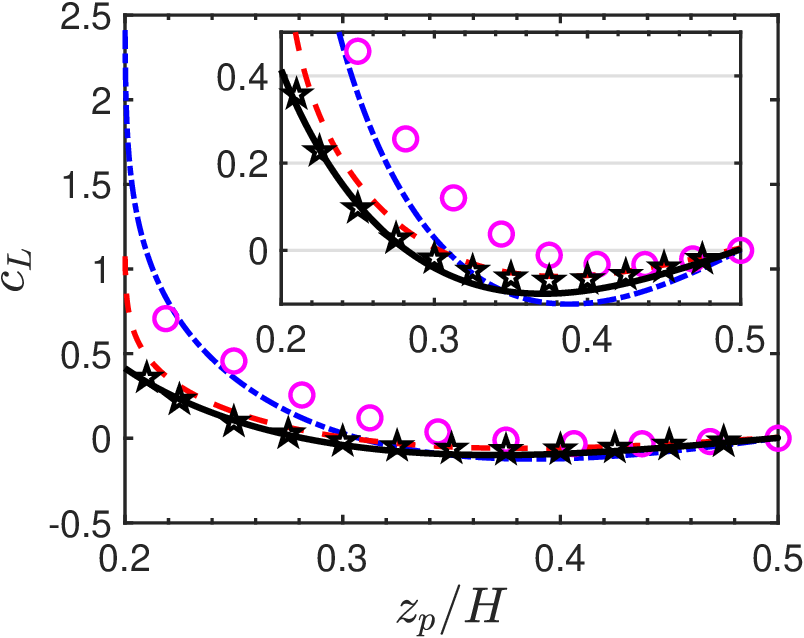}
        \caption*{(b)}
        \label{fig:cL_aH02_model}
    \end{minipage}

    \begin{minipage}{0.48\linewidth}
        \centering
        \includegraphics[trim={0cm 0cm 0cm 0cm}, clip,width=1\textwidth]{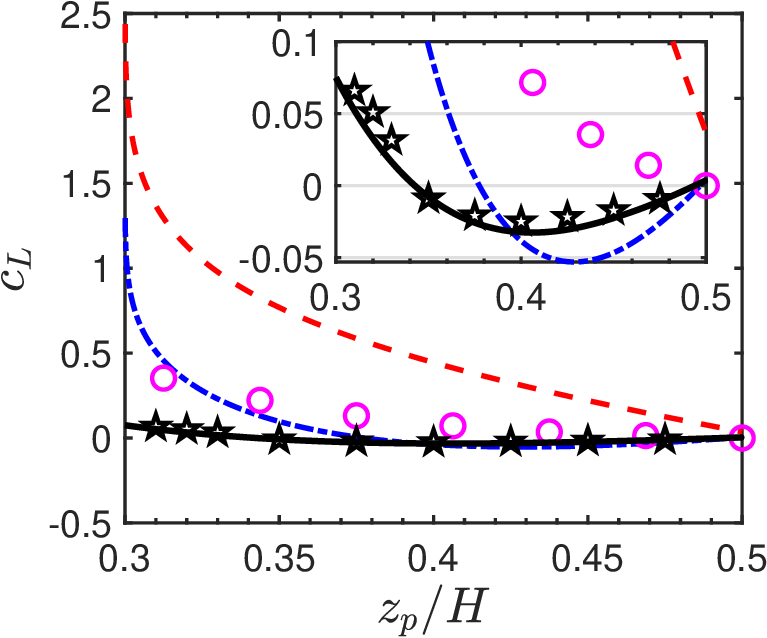}
        \caption*{(c)}
        \label{fig:cL_aH03_model}
    \end{minipage}
    \hfill
    \begin{minipage}{0.49\linewidth}
        \centering
        \includegraphics[trim={0cm 0cm 0cm 0cm}, clip,width=1\textwidth]{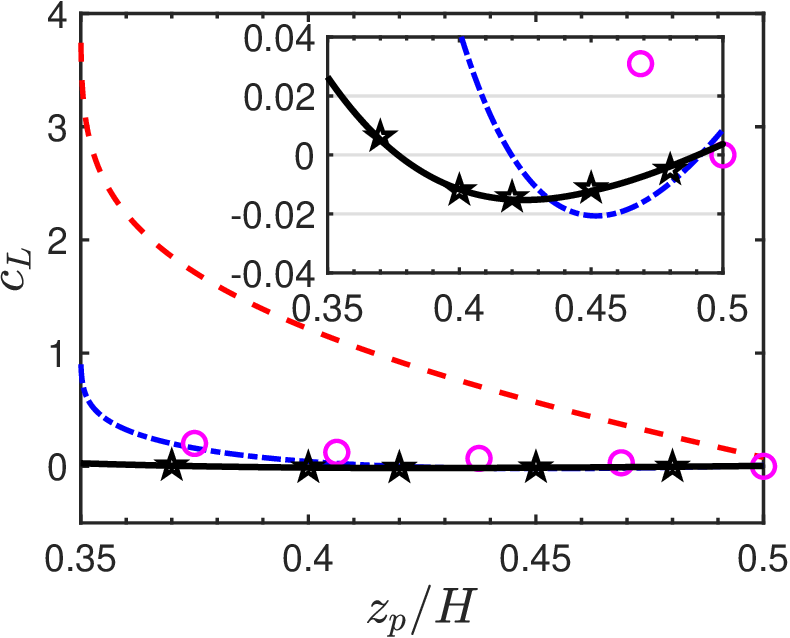}
        \caption*{(d)}
        \label{fig:cL_aH035_model}
    \end{minipage}

\caption{Lift coefficient \(c_L\) as a function of the particle center-to-wall distance \(z_p\) for various particle sizes: (a) \(a/H = 0.1\); (b) \(a/H = 0.2\); (c) \(a/H = 0.3\); (d) \(a/H = 0.35\). Comparisons are made between the present IBM results, the current model predictions based on Eqs.~(\ref{eq:cl_Hood}) and (\ref{eq:c_N}), and published model predictions \cite{asmolov2018inertial, nizkaya2020inertial, hood2015inertial}. {Insets provide magnified views highlighting local differences between numerical and model results, particularly where $c_L$ is small.}}
\label{fig:cL_model_compare}
\end{figure}

As illustrated in Figs.~\ref{fig:cL_model_compare}(a-d), we compare our model's predictions with numerical simulations and existing models \cite{asmolov2018inertial, nizkaya2020inertial, hood2015inertial} for particle sizes \( a/H = 0.1 \), 0.2, 0.3, and 0.35.\
Fig.~\ref{fig:cL_model_compare}(a) presents results for \( a/H = 0.1 \). When the particle is positioned away from the wall (\( z_p/H \ge 0.2 \)), all models align closely with the numerical data. However, near the wall (\( 0.11 \le z_p/H < 0.2 \)), our model exhibits superior agreement.\
For \( a/H = 0.2 \) (Fig.~\ref{fig:cL_model_compare}(b)), the model by \citet{nizkaya2020inertial} provides the most accurate predictions but overestimates the lift force very close to the wall. In contrast, the models of \citet{asmolov2018inertial} and \citet{hood2015inertial} capture the qualitative trend yet deviate quantitatively.\
Larger particles (\( a/H = 0.3 \) and 0.35, Figs.~\ref{fig:cL_model_compare}(c,d)) reveal notable discrepancies: the \citet{nizkaya2020inertial} model significantly overpredicts the lift force, whereas our model maintains consistency with simulations. The frameworks by \citet{asmolov2018inertial} and \citet{hood2015inertial} follow the expected behavior but with reduced accuracy.\
Overall, our model demonstrates robust agreement across all particle sizes and throughout the channel, outperforming prior approaches especially in proximity to the wall.\

\begin{figure}[h]
\centering
\includegraphics[trim={0cm 0cm 0cm 0cm}, clip,width=0.4\textwidth]{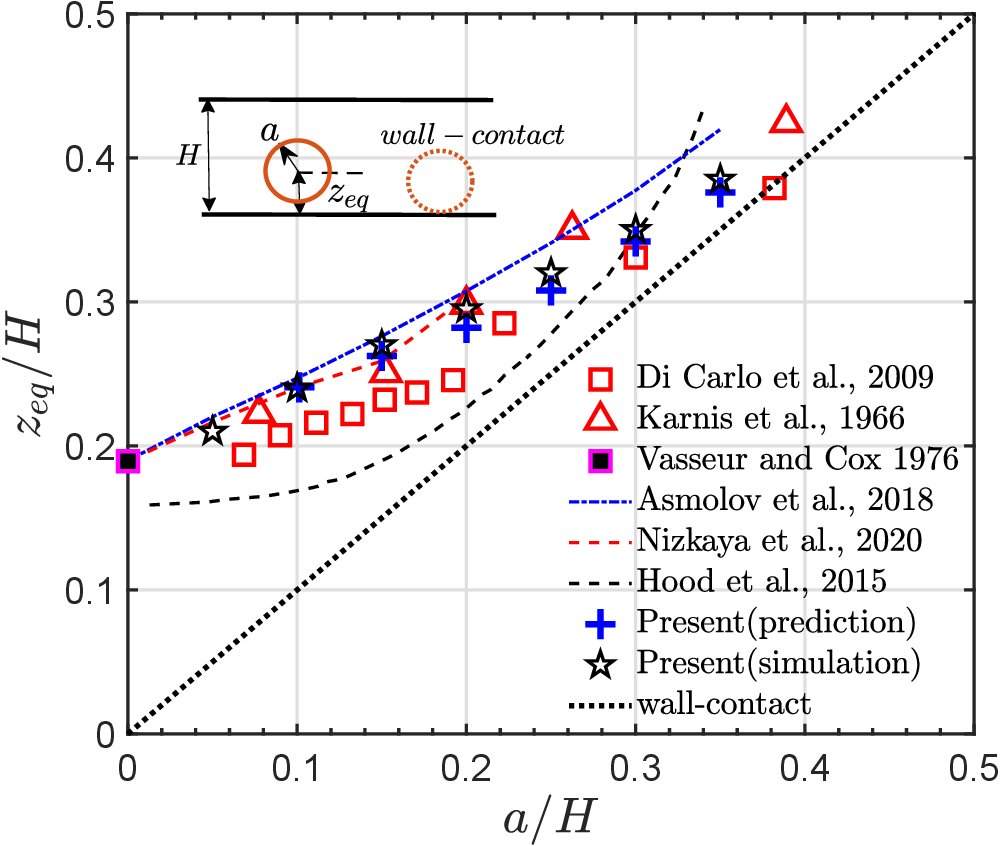}
\caption{Equilibrium positions as a function of particle size. Present model prediction and numerical simulations are for $0.05 \le a/H \le 0.35$. Data from \citet{di2009particle} and \citet{hood2015inertial} correspond to a square channel on symmetry planes, \citet{Karnis1966} corresponds to a tube, and \citet{vasseur1976lateral} is for \(a/H \ll 1\). Model predictions of \citet{asmolov2018inertial} and \citet{nizkaya2020inertial} are for 2D Poiseuille flow.}
\label{fig:Equilirbium_z}
\end{figure}

In microfluidic systems, particle equilibrium states are often the primary experimental outcome, direct measurements of the lift force itself—which governs particle migration—are rarely attainable in practice. Here, as shown in Fig.~\ref{fig:Equilirbium_z}, we use numerical simulations and model predictions, alongside selected published data, to examine the equilibrium position for particle sizes in the range \( 0.05 \leq a/H \leq 0.35 \). The present model, based on Eqs.~(\ref{eq:cl_Hood}) and (\ref{eq:cl_present}), shows good agreement with the numerical results. The equilibrium position, \( z_{eq} \), decreases almost linearly with particle size \( a/H \), trending towards the asymptotic behavior predicted for pointwise particles by \citet{vasseur1976lateral}. Both the values of \( z_{eq} \) and their trends align with the experimental results from \citet{Karnis1966} for particles in a tube flow.

Compared to particles in confined square channels \cite{di2009particle}, the present results reveal a larger particle-wall distance in 2D Poiseuille flow, which becomes more pronounced for \( a/H \leq 0.2 \). The nearest gap between the particle and the wall decreases continuously as the particle size increases. This behavior contrasts with \citet{hood2015inertial}, who predicted a nonmonotonic relationship between the nearest gap and particle size, with the smallest gap occurring at \( a/H \approx 0.22 \). 

Although the model from \citet{asmolov2018inertial} is only valid for \( a/H \leq 0.05 \), it provides reasonable equilibrium position predictions even beyond its intended range of applicability. The model from \citet{nizkaya2020inertial} predicts \( z_{eq} \) accurately up to \( a/H \sim 0.2 \), but for larger particles, it consistently predicts a positive lift force across the channel, limiting its reliability for \( a/H > 0.2 \). Similarly, the present model also predicts a positive lift force across the channel for \( a/H \geq 0.4 \).\

\subsection{\label{sec:3.2 Re_p Effect} Effect of $Re_p$ on the validity of the lift force model}

In the mixed scaling law model proposed by \citet{hood2015inertial}, the flow velocity is expanded in powers of \(Re_p = (a/H)^2 Re\), with the expansion truncated at the first-order term, while terms of order \(Re_p^2\) and higher are neglected. This assumption holds valid for \(Re_p \ll 1\). In our simulations shown in Sec.\ref{sec:3.1 particle size} we present particle sizes in the range \(0.1 \leq a/H \leq 0.35\) for \(Re = 1\), corresponding to \(0.01 \leq Re_p \leq 0.1225\), which falls within the model's valid region. To investigate the effect of \(Re_p\), we examine the lift force coefficient for a larger particle size (\(a/H = 0.3\)) at \(Re = 0.1, 1, 10\), corresponding to \(Re_p = 0.009, 0.09, 0.9\). \

\begin{figure}[h]
\centering
\includegraphics[trim={0cm 0cm 0cm 0cm}, clip,width=0.4\textwidth]{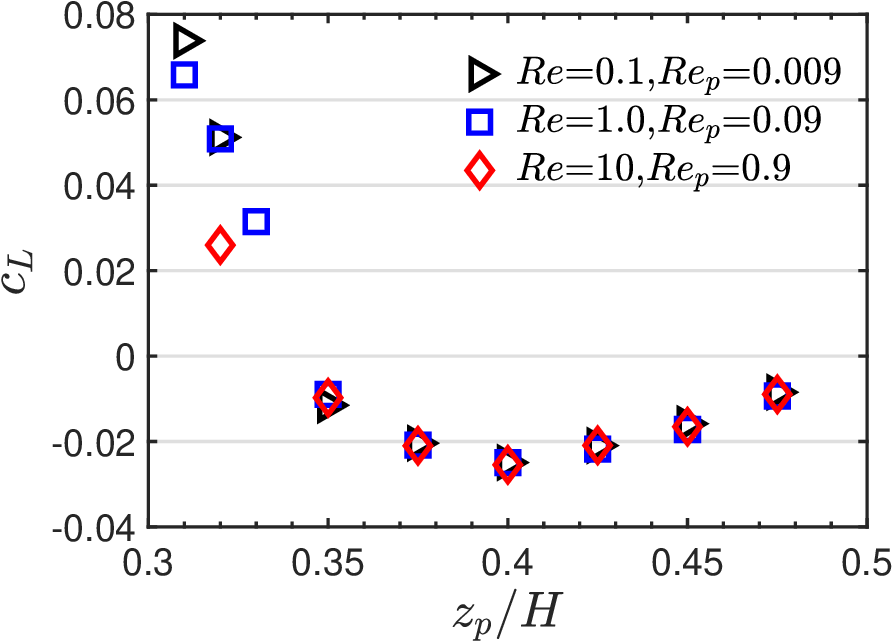}
\caption{Effect of Reynolds number on the lift coefficient \(c_L\) for a large particle (\(a/H = 0.3\)) at different locations.}
\label{fig:aH03_Re}
\end{figure}

As shown in Fig.~\ref{fig:aH03_Re}, the lift coefficient is independent of \(Re_p\) near the channel center across the tested range. However, closer to the channel wall, the lift coefficient decreases as \(Re_p\) increases, with a particularly significant reduction observed at \(Re_p = 0.9\) compared to \(Re_p = 0.09\). This indicates that the model is not valid for \(Re_p \ge 0.9\) when the particle is very close to the wall \( \left(z_p-a\right)/H < 0.05\).

\subsection{\label{sec:3.3 slip wall} Effect of slip wall on the validity of the lift force model}

When fluid flows over superhydrophobic surfaces, slip can occur \cite{quere2008wetting}. In microfluidic devices advances in microfabrication techniques have enabled the development of textured surfaces with nanostructures and specialized coatings, resulting in superhydrophobicity \cite{vrancken2018situ}. The presence of a slip wall can significantly modify the equilibrium position of particles \cite{nizkaya2020inertial_slip}. To account for this, we analyze the lift force on a large particle by applying the Navier boundary condition at the bottom wall, which assumes that the tangential force per unit area exerted on the solid surface is proportional to the slip velocity \cite{vinogradova2011wetting},

\begin{equation}
{\bm{U}_{slip}}=b \left.   \frac{\partial \bm{U(z)}}{\partial{z}}   \right\rvert _{z=0}.
\label{eq:us_Navier}
\end{equation}

\noindent Here, {$\bm{U}_{slip}$} represents the slip velocity at the wall,  ${\partial \bm{U(z)}}/{\partial{z}}$ is the local shear rate, and \(b\) denotes the slip length. We assume isotropic textures, so \(b\) is identical in all three directions. The standard no-slip boundary condition corresponds to \(b = 0\), while the shear-free boundary condition corresponds to \(b \rightarrow \infty\) \cite{vinogradova1995drainage}. In our simulations, we apply slip boundary to the bottom wall while the top wall remains no-slip, we adjust \(b\) to the bottom wall and the pressure gradient in the streamwise direction to achieve streamwise slip velocities of \(U_{slip}/U_m = 0.06\) and \(0.11\) while maintaining \(Re \simeq 1\). For \(Re \lesssim 20\), \citet{nizkaya2020inertial_slip} found that the streamwise velocity profile remains approximately parabolic,
\begin{equation}
\frac{U(z)}{U_m} \simeq 4 \frac{z}{H} \left( 1-\frac{z}{H} \right) +\frac{ U_{slip}}{U_m}\left( 1-\frac{z}{H} \right).
\label{eq:us_Navier}
\end{equation}

\begin{figure}[h]
\centering
\begin{subfigure}{0.35\textwidth}
\includegraphics[trim={0cm 0cm 0cm 0cm}, clip,width=1\textwidth]{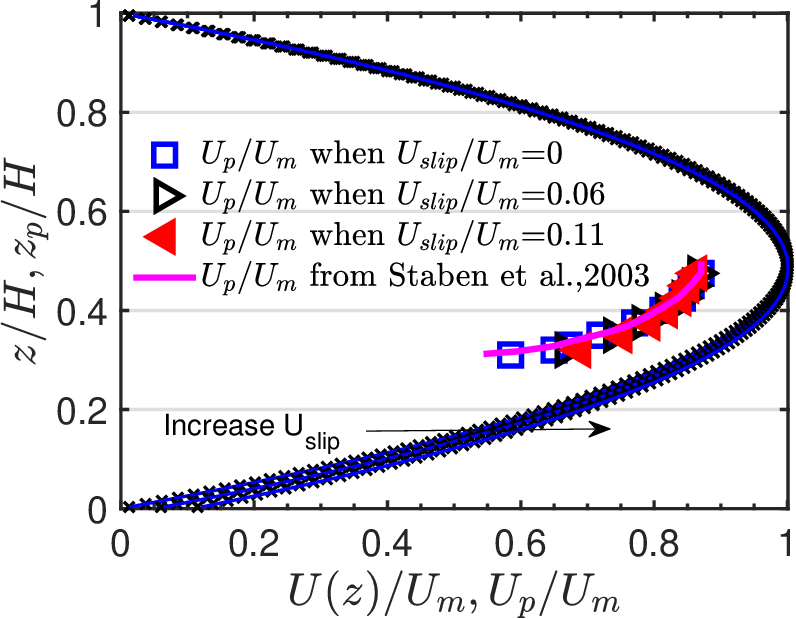}
\caption{}
\label{fig:Slip_velocity}
\end{subfigure}

\begin{subfigure}{0.36\textwidth}
\includegraphics[trim={0cm 0cm 0cm 0cm}, clip,width=1\textwidth]{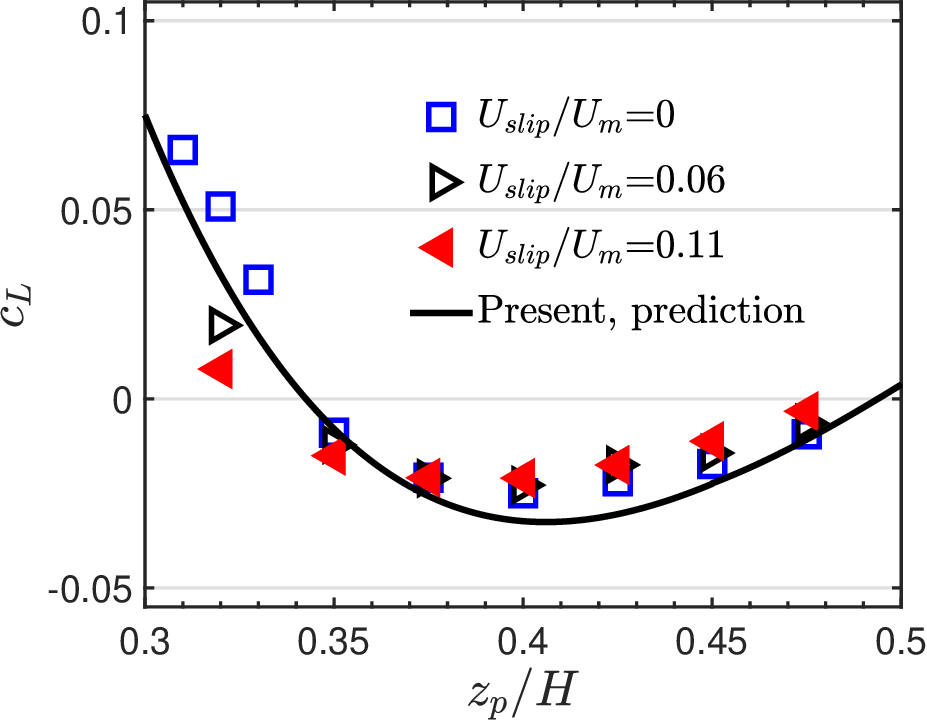}
\caption{}
\label{fig:Slip_cl}
\end{subfigure}

\caption{(a) Comparison of single-phase flow profiles between numerical simulations and predictions from Eq.~(\ref{eq:us_Navier}) for a no-slip wall and slip velocities of \(U_{slip}/U_m = 0.06\) and \(0.11\). Simulation results (cross) and theoretical predictions (solid curves) exhibit close agreement in the plotted profiles. In particle-laden flow, particle migration velocities for different slip velocities are shown and compared with the experimental results from \citet{staben2003motion}. (b) Effect of slip velocity on the lift coefficient \(c_L\) for a large particle (\(a/H = 0.3\)) at different locations, with comparisons to model predictions from Eqs.~(\ref{eq:cl_Hood}) and (\ref{eq:cl_present}).}
\end{figure}

Fig.~(\ref{fig:Slip_velocity}) illustrates that the velocity profile from the present simulation agrees well with the prediction in Eq.~(\ref{eq:us_Navier}). The slip wall primarily influences the bottom half of the velocity profile, while the top half remains largely unaffected. A large particle (\(a/H = 0.3\)) was placed at various locations for \(Re = 1\), and the lift force coefficient is presented in Fig.~(\ref{fig:Slip_cl}). The results indicate a slight increase in the lift force near the channel center and a significant decrease near the wall.
The present model remains valid for predicting a reasonable lift force when \(U_{slip}/U_m \leq 0.06\). Additionally, the equilibrium position moves closer to the slip wall as the slip velocity increases, consistent with the observations of \citet{nizkaya2020inertial_slip} for \(a/H = 0.15\).
While \citet{nizkaya2020inertial_slip} suggest that the reduction in lift force may result from a lower particle slip velocity due to wall slip, our findings reveal that the difference between the particle velocity and fluid velocity changes minimally with increasing wall slip velocity. For instance, at \(z_p/H = 0.32\), the particle slip velocity is \((U_p-U(z_p))/U_m = -0.223, -0.218, -0.210\), corresponding to \(U_{slip}/U_m = 0, 0.06, 0.11\), as shown in Fig.~(\ref{fig:Slip_velocity}).
A more detailed investigation is required to fully understand the mechanisms behind the reduction in lift force.\\

\section{\label{sec:4.Conclusion} Concluding remarks}

In this study, we investigated the lift force acting on a finite-size, neutrally buoyant rigid spherical particle in shallow channel flow at low Reynolds numbers (\(Re \leq 10\)). Using a high-fidelity immersed boundary method, we computed the lift force for particle size ratios in the range \(0.03 \leq a/H \leq 0.35\), where \(a\) is the particle radius and \(H\) is the channel height. To predict the lift force coefficient, we propose an explicit formula for the \(a^4\) and \(a^5\) coefficients from \citet{hood2015inertial} by leveraging lift force models from \citet{asmolov2018inertial} and \citet{nizkaya2020inertial}. Our approach accurately predicts the lift force up to \(a/H = 0.35\), as validated against numerical results.\  

The asymptotic theory is derived using a regular perturbation expansion in the particle Reynolds number (\(Re_p\)), which limits its applicability to \(Re_p \ll 1\). To assess the broader validity of this framework, we examined the effect of \(Re_p\) and found that the proposed model remains effective for \(Re_p \leq 1\). Importantly, our results reveal a significant reduction in the lift force coefficient with increasing \(Re_p\), particularly when the particle is near the wall \( \left(z_p-a\right)/H \le 0.05\).  \

We also explored the influence of slip boundary conditions \cite{nizkaya2020inertial_slip}, motivated by advancements in microfabrication and surface coating technologies that enable surface hydrophobicity. Our simulations show that increasing the slip length reduces the near-wall lift force, shifting the equilibrium position closer to the wall and enhancing particle-wall interactions. For small slip velocities (\(U_{slip}/U_m \leq 0.06\)), the lift force model remains effective in providing reasonable predictions. \

\begin{figure}[h]
    \centering
    \begin{minipage}{0.49\linewidth}
        \centering
        \includegraphics[trim={0cm 0cm 0cm 0cm}, clip,width=1\textwidth]{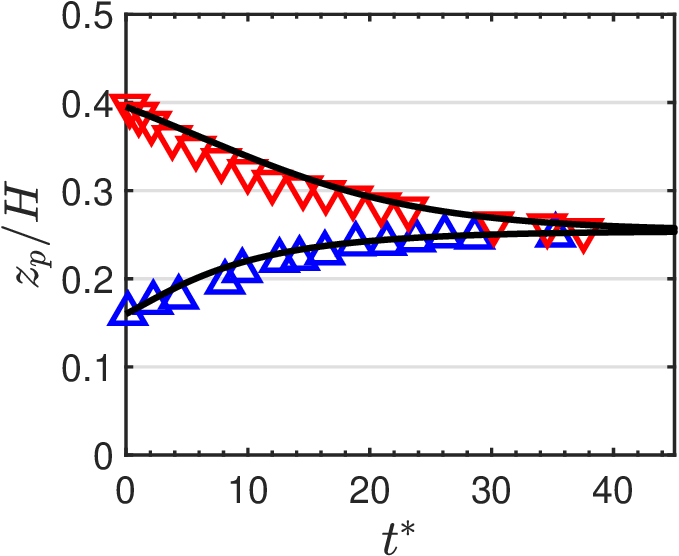}
        \caption*{(a)}
        \label{fig:Trajectory_aH015}
    \end{minipage}
    \begin{minipage}{0.49\linewidth}
        \centering
        \includegraphics[trim={0cm 0cm 0cm 0cm}, clip,width=1\textwidth]{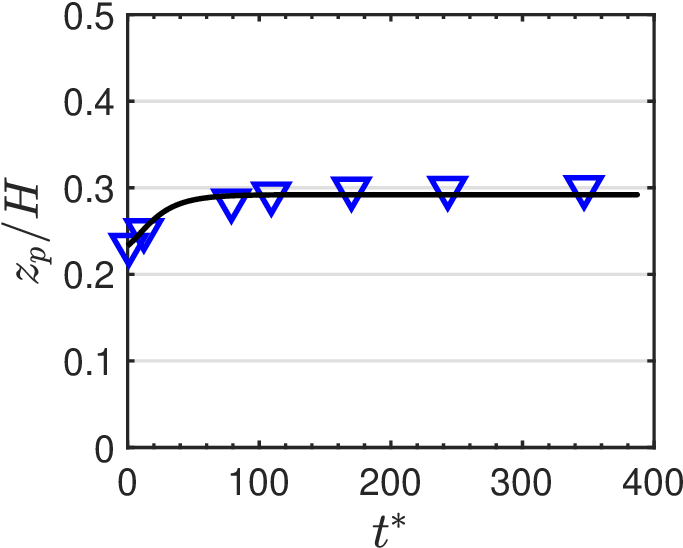}
        \caption*{(b)}
        \label{fig:Trajectory_aH02}
    \end{minipage}
\caption{Comparison of the simulated Segr\'e-Silberberg effect with the experimental results of \citet{Karnis1966}. The channel center is located at \(z_p = 0.5\). Following \citet{feng1994direct}, time is nondimensionalized as \(t^* = \frac{\rho_f \left< U \right> ^2}{\mu} \left(\frac{a}{R}\right)^2 t\). (a) For \citet{Karnis1966}, the particle radius-to-tube diameter ratio is 0.1525, while in our simulation \(a/H = 0.165\), both upward and inward migrations are compared; (b) For \citet{Karnis1966}, the particle radius-to-tube diameter ratio is 0.2, while in our simulation \(a/H = 0.22\), only inward migration is compared.}
\label{fig:Predict_trajectory}
\end{figure}

This simple and practical approach enables rapid estimation of equilibrium positions and particle migration dynamics under different conditions, making it valuable for microfluidic applications. The particle’s streamwise velocity, \(U_p(z_p)\), is estimated using the model from \citet{pasol2011motion,pasol2013corrigendum}, while the wall-normal migration velocity, \(W_p(z_p)\), is determined by balancing the lift force and drag force (Appendix~\ref{app:Stokes_correction}) with corrections to Stokes’ law. Assuming quasi-steady motion in the low Reynolds number regime, the particle trajectory (\(x_p, z_p\)) can be obtained by integrating \(dx_p/dt = U_p\) and \(dz_p/dt = W_p\) from a given initial position. To evaluate the model’s predictive capability, we compared its trajectory predictions with experimental data \cite{Karnis1966} as shown in Fig.~\ref{fig:Predict_trajectory}. The particle size in the model was adjusted to match equilibrium positions with the same initial wall-normal location. {Typically the experimental data used for comparison were obtained in circular tubes or square channels, whereas the present simulations consider a two‑dimensional planar Poiseuille flow. These geometric differences can lead to quantitative discrepancies in lift force and equilibrium position, particularly due to spanwise confinement and wall‑curvature effects that are not captured in the present formulation. Nonetheless, in straight shallow microchannels with large aspect ratios and low particle Reynolds numbers, particle migration is predominantly governed by wall‑normal inertial lift, for which the planar Poiseuille approximation provides a physically transparent and practically useful description. Accordingly, the present comparisons focus on trends and equilibrium behavior rather than exact quantitative agreement.} Consistent with this interpretation, despite geometric differences, \citet{feng1994direct} demonstrated that the behavior of a sphere in a tube closely resembles that of a cylinder in two‑dimensional Poiseuille flow, supporting the relevance of the present comparison. \

While this study focused on neutrally buoyant particles at low \(Re_p\) in shallow channels, several open questions remain. {In this study we assume neutrally buoyant particles in order to isolate inertial lift mechanisms. In experimental systems, however, small density mismatches or particle weight may introduce an additional body force, which can become comparable to inertial lift for larger particles or near equilibrium positions. Such effects may shift the equilibrium location toward and introduce quantitative deviations when comparing with experiments.} One promising direction is to explore the influence of particle inertia on the validity of the lift force model for \(a/H \geq 0.15\), particularly for non-neutrally buoyant particles subjected to gravity. Additionally, further investigation is needed to understand the mechanisms behind lift force reduction due to slip boundary conditions. Another potential research direction is to investigate how external forces—such as dielectric, electrostatic, or acoustic forces—affect the applicability and accuracy of the lift force model in multi-physics systems. Expanding on these aspects would provide deeper insights into particle dynamics in microfluidic devices and broaden the potential applications of the proposed model. 

\begin{acknowledgments}
The authors declare no acknowledgments.
\end{acknowledgments}

\appendix

\section{\label{app:LiftModel} Lift force coefficients proposed by \citet{asmolov2018inertial}, \citet{nizkaya2020inertial}, and \citet{hood2015inertial}}

For small particle size $a/H \le 0.05$, The lift coefficient proposed by \citet{asmolov2018inertial} is defined as 

\begin{equation}
c_{L}=c_{0,V_s}+\gamma c_{1,V_s} V_s+c_{2,V_s} V_s^2,
\label{eq:cl_asmolov}
\end{equation}
\noindent where $\gamma=G(z)/G_m=1-2z_p/H \le 1$ is a dimensionless local shear rate at the particle position. $c_{0,V_s}$, $c_{1,V_s}$, $c_{2,V_s}$ are fitted values listed as follows
\begin{eqnarray}
c_{0,V_s}&=&2.25(\frac{z_p}{H}-0.5)-23.4(\frac{z_p}{H}-0.5)^3, \nonumber \\
c_{1,V_s}&=&-3.24139 \zeta -2.676 \nonumber \\
& &-0.8248 \zeta^{-1}+0.4616 \zeta^{-2}, \nonumber\\
c_{2,V_s}&=&1.7631+0.3561 \zeta^{-1} \nonumber\\
& &-1.1837 \zeta^{-2}+0.845163 \zeta^{-3},\nonumber\\
\label{eq:c0-2_asmolov}
\end{eqnarray}
\noindent where $\zeta=z_p/a$, $c_{0,V_s}$ is from \citet{vasseur1976lateral} for pointwise particle, and $c_{1,V_s}$ and $c_{2,V_s}$ are from \citet{cherukat1994inertial} for finite-size particle in a near-wall shear flow.

In Eq.~(\ref{eq:cl_asmolov}) the slip velocity $V_s$ is based on linear shear flow which can be represented by a correction function $h$ from \citet{wakiya1967particle} and \citet{reschiglian2000standardless}

\begin{equation}
V_s=\frac{z(H-z)(h-1)}{aH},
\label{eq:Slip_asmolov}
\end{equation}

\noindent with 
\begin{eqnarray}
h&=&\frac{200.9b-(115.7b+721)\zeta^{-1}-781.1}{-27.25b^2+398.4b-1182}, \zeta < 3,\nonumber \\
h&=&\frac{1-\frac{5}{4}\zeta ^{-3}+\frac{5}{4}\zeta ^{-5}-\frac{23}{48}\zeta ^{-7}-\frac{1375}{1024}\zeta ^{-8}}{1-\frac{15}{16}\zeta ^{-3}+\zeta ^{-5}-\frac{3}{8}\zeta ^{-7}-\frac{4565}{4096}\zeta ^{-8}}, ~\zeta \geq 3, \nonumber \\
\label{eq:h_asmolov}
\end{eqnarray}
\noindent where $b=\ln(\zeta-1)$.\\

For medium particle size $a/H \le 0.15$, \citet{nizkaya2020inertial} proposed a modification of expression for the lift coefficient

\begin{equation}
c_{L}=c_{0,V{^\prime}_s}+c_{1,V{^\prime}_s} V^{\prime}_s+c_{2,V{^\prime}_s} {V{^\prime}_s}^2,
\label{eq:cl_Nizkaya}
\end{equation}
\noindent where $c_{0,V{^\prime}_s}$ and $c_{1,V{^\prime}_s}$ are modifications of $c_{0,V_s}$ and $c_{1,V_s}$ in Eqs.(\ref{eq:c0-2_asmolov}), 

\begin{eqnarray}
c_{0,V{^\prime}_s}&=&\beta_0 c_{0,V_s}, \nonumber \\
c_{1,V{^\prime}_s}&=&4\left(1-\frac{2z_p}{H} \right) \beta_1 c_{1,V_s}.
\label{eq:c_N}
\end{eqnarray}

In Eqs.(\ref{eq:c_N}), $\beta_0$ and $\beta_1$ are correction factors based on simulation data,

\begin{eqnarray}
\beta_0&=&1+3.32\frac{a}{H}-26.45\left(\frac{a}{H} \right)^2, \nonumber \\
\beta_1&=&1-8.39\frac{a}{H}+19.65\left(\frac{a}{H} \right)^2.
\label{eq:beta_Nizkaya}
\end{eqnarray}

The coefficient $c_{2,V{^\prime}_s}$ is given by \citet{Cherukat_Mclaughlin_1995}
\begin{eqnarray}
c_{2,V{^\prime}_s}&=&1.8065+0.89934 \zeta^{-1} \nonumber\\
& &-1.961 \zeta^{-2}+1.02161 \zeta^{-3}.
\label{eq:c2_Nizkaya}
\end{eqnarray}

Additionally, the Faxen correction due to the parabolic flow profile is included to predict the particle slip velocity

\begin{equation}
V^\prime_s=\frac{U(z_p)(h-1)-4/3\left(a/H\right)^2}{U_m},
\label{eq:Vs_Nizkaya}
\end{equation}
\noindent where the particle velocity correction coefficient $h$ is the same as given by Eq.(\ref{eq:h_asmolov}).\\


{
For arbitrary particle sizes $0< a/H < 0.5$, \citet{hood2015inertial} proposed a perturbation expansion of the lift force in powers of the particle Reynolds number, leading to the mixed scaling

\begin{equation}
c_L=c_4+c_5*(a/H)
\label{eq:cL_Hood}
\end{equation}

where the coefficients $c_4$ and $c_5$ depend solely on the particle wall‑normal position ($z_p/H$) and are independent of particle size. These coefficients are not given in closed analytical form but are determined numerically by solving the Stokes and inertial correction problems associated with a finite‑size sphere in Poiseuille flow. Discrete values of $c_4$ and $c_5$ for selected particle locations are reported in the supplementary material of \citet{hood2015inertial} for three‑dimensional square channels. In the present work, the coefficient values on the symmetry plane are used as representative inputs for comparison with two‑dimensional planar Poiseuille flow in Fig.\ref{fig:cL_validation} and Fig.\ref{fig:cL_model_compare}.
} 

\section{\label{app:Stokes_correction}The Stokes drag correction to predict migration velocity confined by two walls}

When a rigid sphere of radius \(a\) moves with a constant velocity \(W_p\) toward a wall in the low particle Reynolds number regime, \citet{brenner1961slow} solved the creeping motion equations and theoretically predicted the correction to Stokes' law due to the presence of the solid wall:

\begin{equation}
F_d=- 6 \pi \mu a W_p \lambda,
\label{eq:Fd_Brenner}
\end{equation}

\noindent where \(\lambda\) is the correction coefficient given by:

\begin{eqnarray}
\lambda&=& \frac{4}{3} \sinh (\alpha) \sum_{n=1}^{\infty} \frac{n(n+1)}{(2n-1)(2n+3)}  \nonumber\\ 
& & \left( \frac{2\sinh [(2n+1) \alpha] +(2n+1)\sinh (2\alpha) }{4\sinh^2 \left[ (n+\frac{1}{2})\alpha \right] -(2n+1)^2 \sinh^2 (\alpha)}-1\right), \nonumber\\ 
\label{eq:lambda_Brenner}
\end{eqnarray}

\noindent where \( \alpha=\cosh ^{-1} (1+\epsilon) \) and \(\epsilon = (z_p - a)/a\). The complex function for \(\lambda\) above was simplified by \citet{smart1989measurement} as:

\begin{equation}
\lambda=1+\frac{1}{\epsilon}. \nonumber
\end{equation}

When considering the rigid sphere, located at \(z_p\), moves toward the bottom wall (\(z=0\)) in the presence of a top wall (\(z=H\)), \citet{asmolov2018inertial} extended the approach by superimposing the drag corrections from both walls. In the low Reynolds number regime, this approach leverages the linearity of the Stokes equations to account for additive effects. The resulting correction coefficient is:

\begin{equation}
\lambda=1+\frac{1}{\epsilon}+\frac{1}{H/a-2-\epsilon}.
\label{eq:lambda_Asmolov}
\end{equation}

Assuming the particle motion is quasi-stationary in the low Reynolds number regime, the lateral migration velocity of the particle can be predicted by balancing the drag force with the lift force \(F_L+F_d=0\), such that:

\begin{equation}
W_p=F_L/(6 \pi \mu a \lambda),
\label{eq:Wp_FL}
\end{equation}

\noindent where ($\lambda$) is given by Eq.~(\ref{eq:lambda_Asmolov}). 

\begin{figure}[h]
\centering
\includegraphics[trim={0cm 0cm 0cm 0cm}, clip,width=0.43\textwidth]{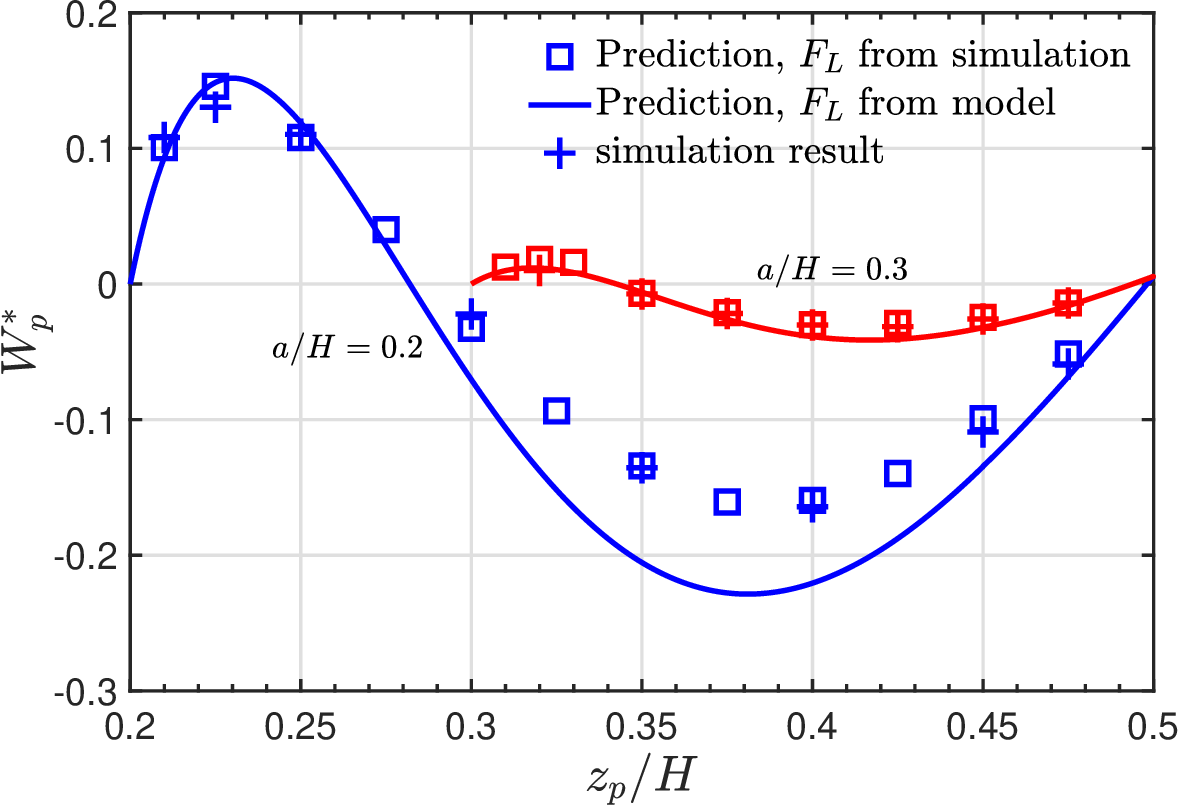}
\caption{Comparison of particle lateral migration velocities for \( a/H = 0.2, 0.3 \) between numerical simulations and model predictions. In the model prediction, \( W_p \) is obtained by Eq.~\ref{eq:Wp_FL}. Both numerically simulated and predicted \( F_L \) values are presented. \( W_p^* \) is the dimensionless form of \( W_p \), defined as \( W_p^* = \left( W_p / aG_m \right) \left( 6\pi / Re_p \right) \), as suggested by \citet{gupta2018conditional}.}
\label{fig:Wp_compare}
\end{figure}

We compare the particle lateral migration velocity \(W_p\) obtained from numerical simulations with the model prediction in Eq.~(\ref{eq:Wp_FL}) for large particle size ratio \(a/H = 0.2, 0.3\) at \(Re = 1\), corresponding to \(Re_p = 0.04, 0.09\). To measure \( W_p \), we follow \citet{nizkaya2020inertial} by fixing the particle in the wall-normal direction while allowing it to rotate and move freely in all other directions. The results, shown in Fig.~\ref{fig:Wp_compare}, indicate that the model prediction based on correction of Stokes equation in Eq.~(\ref{eq:Wp_FL}) gives a good prediction of the particle lateral migration velocity across the entire channel. 

\clearpage
\bibliography{Reference}

\end{document}